\documentclass[12pt]{article}
\pdfoutput=1
\usepackage[utf8]{inputenc}
\usepackage[T1]{fontenc}
\usepackage{float}
\usepackage{authblk}
\usepackage[normalem]{ulem}
\usepackage{amsmath}
\usepackage{amsthm}
\usepackage{amssymb}
\usepackage{amsfonts}
\usepackage{adjustbox}
\usepackage{bbm}
\usepackage{graphicx}
\usepackage{booktabs}
\usepackage{mathtools}
\usepackage{multirow,array}
\usepackage{fullpage}
\usepackage{comment}
\usepackage[mathlines]{lineno}
\usepackage{multirow}
\usepackage{longtable}
\usepackage{subcaption}
\usepackage[table]{xcolor}

\usepackage[style=numeric-comp,sorting=none,maxnames=16]{biblatex}

\usepackage[colorlinks = true,
            linkcolor = teal,
            urlcolor  = teal,
            citecolor = teal]{hyperref}

\title{Spatial scales of COVID-19 transmission in Mexico}

\author[1,2]{Brennan~Klein\thanks{\href{mailto:b.klein@northeastern.edu}{b.klein@northeastern.edu}}}
\author[1]{Harrison~Hartle}
\author[1]{Munik~Shrestha}
\author[3]{Ana~Cecilia~Zenteno}
\author[4]{David~Barros~Sierra~Cordera}
\author[5]{José~R.~Nicolas-Carlock}
\author[6]{Ana~I.~Bento}
\author[7,8]{Benjamin~M.~Althouse}
\author[9,10,11]{Bernardo~Gutierrez}
\author[9,11]{Marina~Escalera-Zamudio}
\author[12,13]{Arturo~Reyes-Sandoval}
\author[9,14,18]{Oliver~G.~Pybus}
\author[1,2]{Alessandro~Vespignani}
\author[15]{Jose~Alberto~Diaz-Quiñonez*\thanks{\href{mailto:alberto_diaz@uaeh.edu.mx}{alberto\_diaz@uaeh.edu.mx}}}
\author[1,16,17]{Samuel~V.~Scarpino*\thanks{\href{mailto:s.scarpino@northeastern.edu}{s.scarpino@northeastern.edu}}}
\author[9,18]{Moritz~U.G.~Kraemer*\thanks{\href{mailto:moritz.kraemer@biology.ox.ac.uk}{moritz.kraemer@biology.ox.ac.uk}}}

\affil[1]{Network Science Institute, Northeastern University, Boston, Massachusetts, USA}
\affil[2]{Laboratory for the Modeling of Biological \& Socio-technical Systems,\protect\\Northeastern University, Boston, Massachusetts, USA}
\affil[3]{Massachusetts General Hospital, Boston, Massachusetts, USA}
\affil[4]{Instituto Mexicano del Seguro Social, Ciudad de México, México}
\affil[5]{Instituto de Investigaciones Jurídicas, Universidad Nacional Autónoma de México,\protect\\Ciudad de México, México}
\affil[6]{Department of Epidemiology and Biostatistics, School of Public Health,\protect\\Indiana University, Bloomington, Indiana, USA}
\affil[7]{Information School, University of Washington, Seattle, Washington, USA}
\affil[8]{Department of Biology, New Mexico State University, Las Cruces, New Mexico, USA}
\affil[9]{Department of Biology, University of Oxford, Oxford, UK}
\affil[10]{School of Biological \& Environmental Sciences,\protect\\Universidad San Francisco de Quito, Quito, Ecuador}
\affil[11]{Consorcio Mexicano de Vigilancia Genómica}
\affil[12]{The Jenner Institute, University of Oxford, Oxford, UK}
\affil[13]{Instituto Politécnico Nacional, IPN, Ciudad de México, México}
\affil[14]{Department of Pathobiology and Population Science, Royal Veterinary College, London, UK}
\affil[15]{Instituto de Ciencias de la Salud, Universidad Autónoma del Estado de Hidalgo,\protect\\Pachuca, Hidalgo, México}
\affil[16]{Institute for Experiential AI, Northeastern University, Boston, Massachusetts, USA}
\affil[17]{Santa Fe Institute, Santa Fe, New Mexico, USA}
\affil[18]{Pandemic Sciences Institute, University of Oxford, UK}

\begin{document}
\maketitle

\clearpage
\begin{abstract}
During outbreaks of emerging infectious diseases, internationally connected cities often experience large and early outbreaks, while rural regions follow after some delay \cite{brockmann_hidden_2013, may_spatial_1984, rice_variation_2021, viboud_synchrony_2006, Davis2021, Balcan21484}. This hierarchical structure of disease spread is influenced primarily by the multiscale structure of human mobility \cite{watts_multiscale_2005, alessandretti_scales_2020, susswein_ignoring_2021}. However, during the COVID-19 epidemic, public health responses typically did not take into consideration the explicit spatial structure of human mobility when designing non-pharmaceutical interventions (NPIs). NPIs were applied primarily at national or regional scales \cite{oliu_barton_green_2021}. Here we use weekly anonymized and aggregated human mobility data and spatially highly resolved data on COVID-19 cases, deaths and hospitalizations at the municipality level in Mexico to investigate how behavioural changes in response to the pandemic have altered the spatial scales of transmission and interventions during its first wave (March - June 2020). We find that the epidemic dynamics in Mexico were initially driven by SARS-CoV-2 exports from Mexico State and Mexico City, where early outbreaks occurred. The mobility network shifted after the implementation of interventions in late March 2020, and the mobility network communities became more disjointed while epidemics in these communities became increasingly synchronised. Our results provide actionable and dynamic insights into how to use network science and epidemiological modelling to inform the spatial scale at which interventions are most impactful in mitigating the spread of COVID-19 and infectious diseases in general.
\end{abstract}

\vspace{2.5cm}
\begin{table}[h]
    \small
    \centering
    \caption{\textbf{Policy summary}}\vspace{-0.35cm}
    {\renewcommand{\arraystretch}{1.5}
    \begin{tabular}{|p{2.7cm}|p{12.8cm}|}
        \hline
        \textbf{Background} & \textit{The establishment, persistence and growth rates of COVID-19 mainly depend on human mobility and mixing. However, current approaches attempting to limit transmission have been primarily based on administrative boundaries instead of the natural scales of human mobility.} \\ \hline
        \textbf{Main findings\newline \& limitations} & \textit{Using aggregated and anonymized human mobility and detailed COVID-19 case data, we find that the scales of human mixing shift during the pandemic and that transmission is highly clustered amongst mobility communities.} \\ \hline
        \textbf{Policy\newline implications} & \textit{Structuring interventions based on spatial mobility may be more effective compared to interventions based on administrative boundaries. Future pandemic control interventions should consider empirical human mobility networks when designing interventions.} \\ \hline
    \end{tabular}}
    \label{tab:policy_summary}
\end{table}

\clearpage
\section{Introduction}

The transmission of infectious diseases is highly heterogeneous. Differences in population structure, the landscape of prior immunity, and environmental factors, result in differences in the timing of outbreaks, their magnitude, and duration \cite{may_spatial_1984, levin_problem_1992, rice_variation_2021, susswein_ignoring_2021, masters_finescale_2020, rader_crowding_2020, lee_spatial_2022, smith_spatial_2005, yang_estimating_2021, rosensteel_characterizing_2021, perkins_heterogeneity_2013, burghardt_unequal_2022, wu_spatial_2022}. For infectious diseases, one principal component determining the spatial structure of outbreaks is the frequency of interactions between susceptible and infectious individuals within and between regions. In most geographies, public health decision-making authority follows political boundaries. However, from an epidemiological perspective, the relevant spatial units may not strictly follow political boundaries but rather human mixing \cite{alessandretti_scales_2020, lee_spatial_2022, nelson_economic_2016}. Evaluating the spatial structure of COVID-19 transmission remains important in determining optimal interventions (non-pharmaceutical and/or vaccination) to reduce transmission and limit the risk of resurgence of cases \cite{graham_measles_2019, lee_engines_2020, edsberg_mollgaard_understanding_2022, DiDomenico2020}.

During the first half of 2020, Mexico experienced one of the largest SARS-CoV-2 epidemics worldwide, with more than 600,000 cases and 65,000 confirmed deaths 
reported between February and September 2020 \cite{coronavirus_mexico} (Fig.~\ref{fig:fig1}a). The epidemic wave peaked in May in the largest metropolitan areas of Mexico City and the State of Mexico and later ignited epidemics in all other states \cite{taboada_genomic_2020}, peaking between June and July 2020 (Fig.~\ref{fig:fig1}b). Here we combine municipality level epidemiological data with weekly anonymized aggregated human mobility data at the same scale, to characterise the spatial scales of the Mexican COVID-19 pandemic and their implications for the implementation of spatially targeted interventions.

\section{Results}\label{sec:results}

\subsection{Spatial expansion of COVID-19 in Mexico}\label{sec:demographics}
\begin{figure}[t!]
    \centering
    \includegraphics[width=1.0\columnwidth]{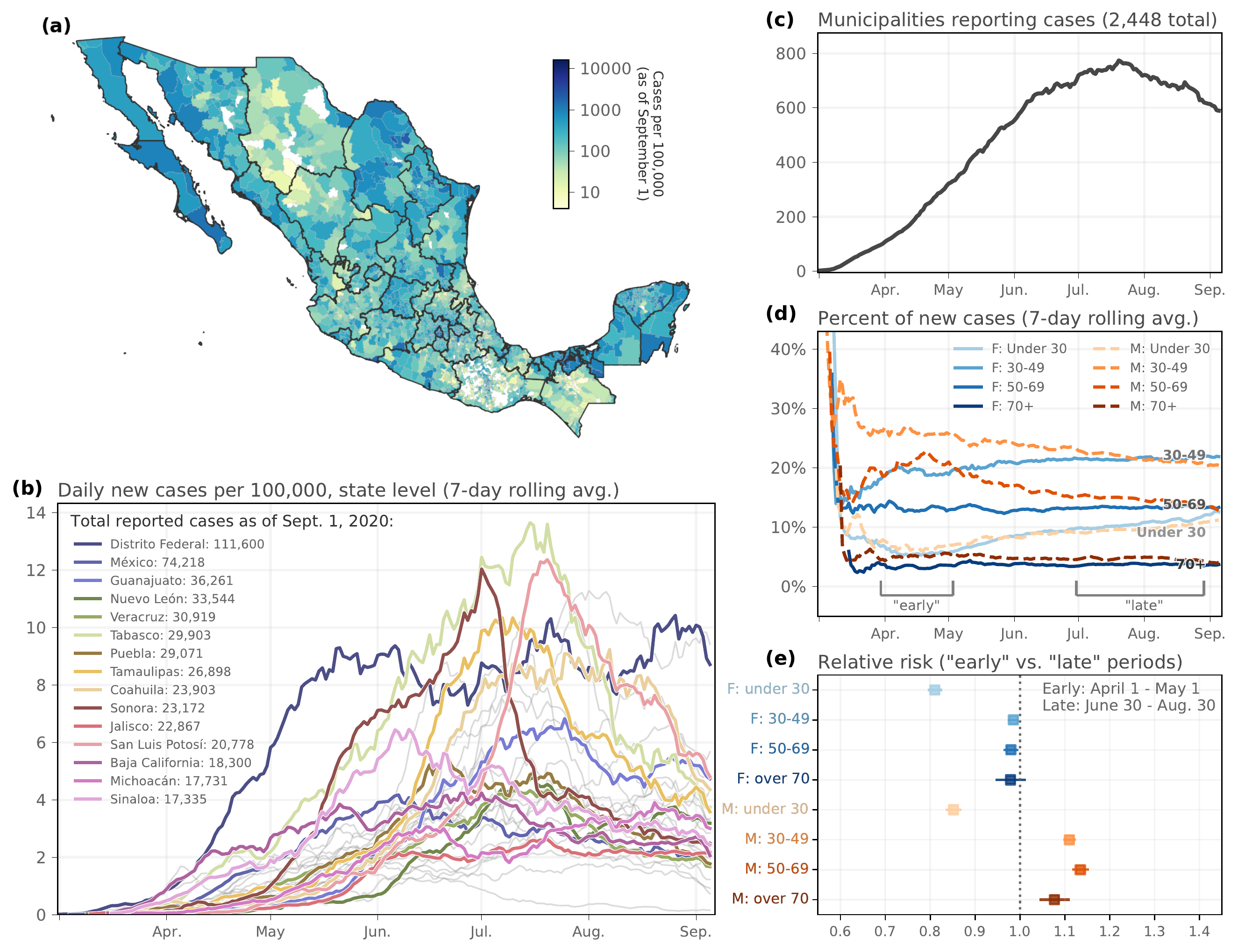}
    \caption{\textbf{Epidemiological situation of COVID-19 in Mexico.} \textbf{(a)} Map of cumulative cases per 100,000 people, as of September 1, 2020. \textbf{(b)} Timeline of new cases per 100,000 population at the state level (7-day rolling average), highlighting the 15 states with the most severe cumulative outbreaks. \textbf{(c)} Number of municipalities that reported confirmed cases of COVID-19 through time. \textbf{(d)} Age and sex distributions of confirmed COVID-19 cases across Mexico, highlighting ``early'' and ``late'' periods during which the relative risk of infections were calculated. \textbf{(e)} Age and sex relative risk ratios of infection, comparing the early vs. late periods from panel \textbf{(d)}.}
    \label{fig:fig1}
\end{figure}

In Mexico, the spatial range of transmission expanded rapidly after reports of the earliest cases in March 2020, with over 700 municipalities reporting transmission by July 2020 (out of 2,448, Fig.~\ref{fig:fig1}c). During April and May the risk of positive RTq-PCR confirmed cases amongst men aged 30-69 was 1.4 times higher than between July 1 and September 1 (Fig.~\ref{fig:fig1}d,e), indicating that the epidemic spread initially within and through these age groups (Extended Data Figure \ref{fig:SI_region_cases}). This dynamic trend in the demographics of cases is similar to that observed in other countries during the early stages of the pandemic \cite{kraemer_effect_2020, monod_age_2021}.

\begin{figure}[t!]
    \centering
    \includegraphics[width=1.0\columnwidth]{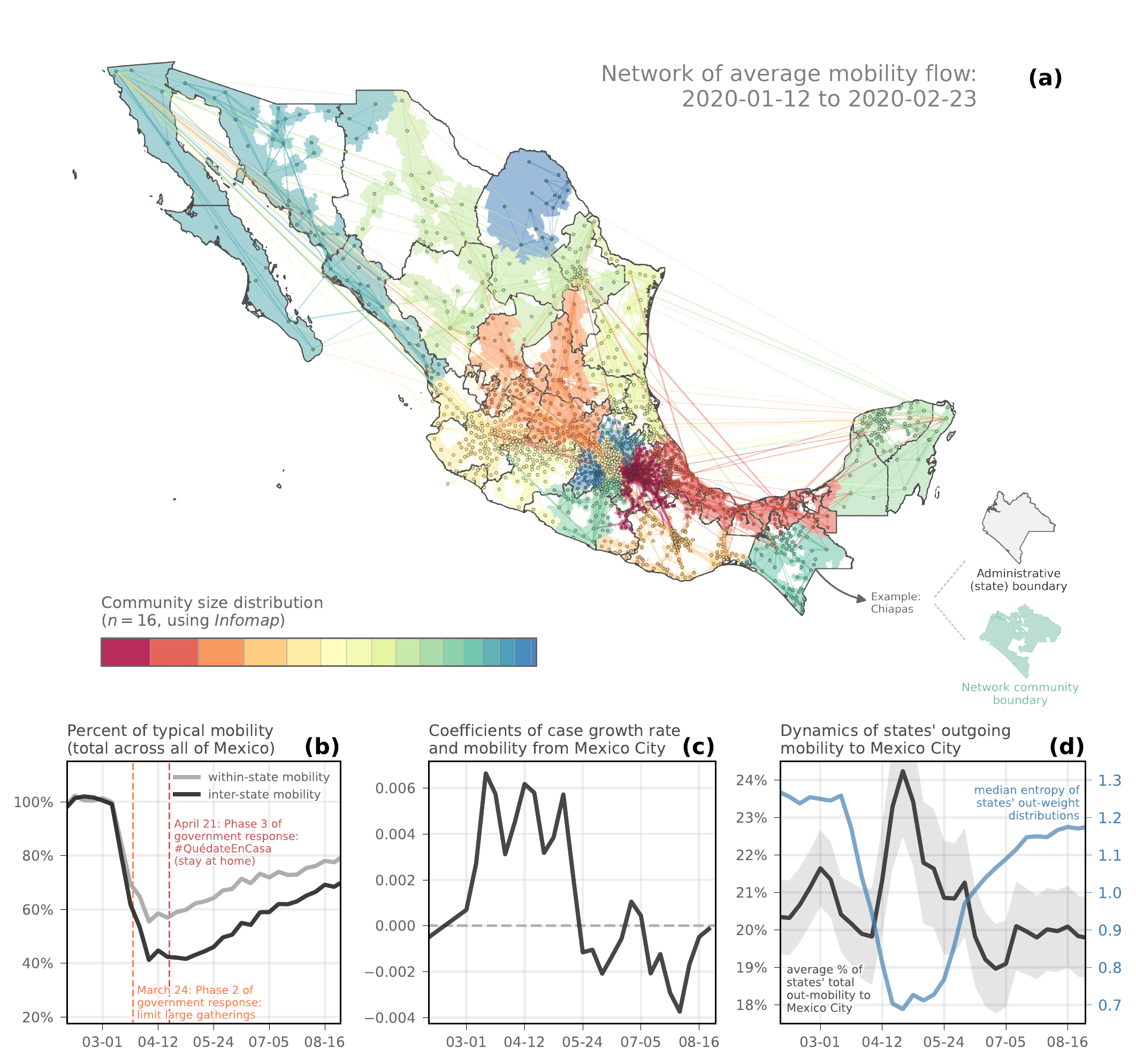}
    \caption{\textbf{Human mobility and transmission of COVID-19 in Mexico.} \textbf{(a)} Pre-pandemic average of the inter-municipality mobility network, coloured by network community (detected using the Infomap algorithm). Mobility flow data is based on the aggregated Google Mobility Research dataset (see Materials \& Methods). \textbf{(b)} Percent of typical weekly mobility nationwide (typical refers to mobility between January 12 and February 29, 2020). \textbf{(c)} Evolution of the coefficients of mobility flow from Mexico City in (lagged) correlations with state-level case rates across the country, highlighting the key role that mobility from Mexico City played in the early stage of the epidemic. \textbf{(d)} Average fraction of total outgoing mobility from each state that is to Mexico City (black) and the median entropy of states' distributions of outgoing mobility. Error bands correspond to 95\% confidence intervals.}
    \label{fig:fig2}
\end{figure}

States that experienced early transmission were the state of Mexico and Mexico City (Fig.~\ref{fig:fig1}b) \cite{taboada_genomic_2020}. Due to the centrality of Mexico City connecting people from abroad (international arrivals) and within Mexico we hypothesise that human mobility from these states was a key driver of the spread of COVID-19 in Mexico. Using anonymized, opt-in and aggregated human movement data from mobile phones (Materials and Methods) we find that case growth rates across Mexican states were well predicted by a lagged model of human movements from the State of Mexico and Mexico City between March and May 2020 (Fig.~\ref{fig:fig2}c, conditional $R^2 = 0.62$; see Materials \& Methods). Further, we observe that the share of overall relative human mobility to and from Mexico and Mexico City increased markedly during that period (Fig.~\ref{fig:fig2}b) when overall human mobility between states declined (Fig.~\ref{fig:fig2}b, Extended Data Figure \ref{fig:SI_withinbetween} showing state level data on change in human mobility). This points towards a change in the network structure of human mobility in Mexico, as documented in some other countries \cite{schlosser_covid19_2020, brown_impact_2021}. Overall transmission, and the importance of Mexico City driving the epidemic, declined after the implementation of NPIs through May 2020. However, after the lifting of physical distancing measures on June 1st (see table of documented changes in NPIs, Table \ref{tab:s1}), case growth rates in the country increased again as a function of mobility from Mexico City, in line with models predicting that lifting lockdowns can lead to reseeding of transmission chains from larger to smaller cities where epidemics were successfully controlled (Fig.~\ref{fig:fig2}b, Table \ref{tab:s1}, \cite{watts_multiscale_2005}).

Variation in weekly new cases within each state in Mexico are generally well predicted by cases in Mexico City weighted by human mobility except for Baja California, Morelos, Chihuahua, Oaxaca, and Chiapas (Extended Data Figure \ref{fig:SI_from_to_mxdf}). We hypothesise that epidemics there were possibly seeded from other countries (USA and Guatemala); further SARS-CoV-2 genomic analyses of unbiased collections of samples will be needed to confirm the SARS-CoV-2 lineage dynamics in these states \cite{taboada_genomic_2020, hill_progress_2021, kraemer_spatiotemporal_2021, du_plessis_establishment_2021, gutierrez_emergence_2022, castelan_sanchez_comparing_2022}. Human mobility data showing cross border (US to Mexico) movements indicate higher overall mobility to bordering states in Mexico and growth rates in US-Mexico border states appear higher in the period between 24 May - 28 June 2020 (Extended Data Figures \ref{fig:SI_border_mobility}, \ref{fig:SI_border_cases}, \ref{fig:SI_border_cases_mobility}). The high mobility during that phase resulted in larger case numbers in states bordering the US when compared to other states in Mexico (Extended Data Figure  \ref{fig:SI_border_cases}).

\begin{figure}[t!]
    \centering
    \includegraphics[width=1.0\columnwidth]{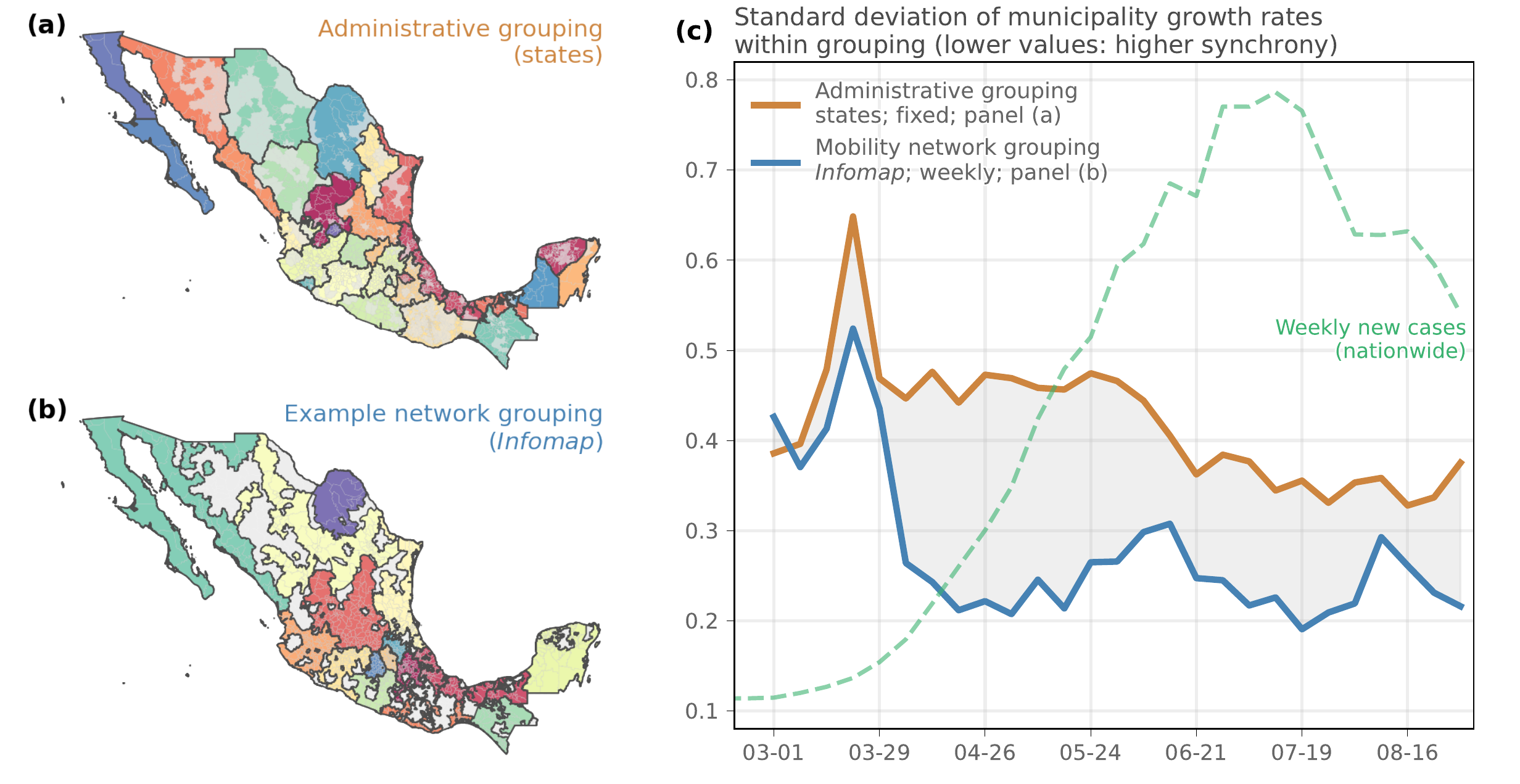}
    \caption{\textbf{Network structure determines the synchrony of epidemics.} \textbf{(a)} Grouping of municipalities based on the state administrative boundaries. Shaded municipalities are removed from downstream analyses as they could not be assigned a movement community (see Materials \& Methods). \textbf{(b)} Example grouping of municipalities based on human movement data and a community detection algorithm \cite{zheng_random_2008} (Materials and Methods). Colours indicate movement communities. Grey municipalities have limited recorded movements and could not be assigned to a community and were consequently excluded from analysis. \textbf{(c)} Synchrony of weekly growth rates of epidemics across municipalities as measured by the pairwise standard error between growth rates. The lower the error, the more synchronised epidemics are. Blue line shows grouping by network communities, and orange shows groupings by state administrative boundaries. The green dashed line shows the nationwide trend in reported cases during this period. For a visual intuition of the differences in within-community standard deviations of growth rates, see Extended Data Figure \ref{fig:SI_intuition}.}
    \label{fig:fig3}
\end{figure}

\subsection{The scales of COVID-19 transmission}\label{sec:results_scales}

It is well known that reductions in mobility (a proxy for reductions in population mixing) have reduced the transmission of COVID-19 within a location \cite{nouvellet_reduction_2021}. However, it remains unclear how structural changes to the mobility network (shifts in the frequency and intensity of mobility within and among regions) have impacted COVID-19 dynamics empirically \cite{schlosser_covid19_2020, fontanelli_intermunicipal_2022, Klein2022, mas_spatiotemporal_2021, brown_impact_2021}. Our underlying hypothesis is that more tightly connected communities exhibit more synchronised epidemic dynamics and, conversely, that more disjointed individual communities have less synchronised epidemics and their epidemics are more likely to fade out \cite{viboud_synchrony_2006, Davis2021, Balcan21484} (here, communities are equivalent to municipalities and synchrony is defined as the similarity among communities in weekly case growth rates \cite{van_panhuis_regionwide_2015}). Both processes have critical implications for disease mitigation and eliminations locally, and at a country level \cite{watts_multiscale_2005, metcalf_rubella_2011, keeling_metapopulation_2004, valdano_use_2022, Chang2022, Nicolau2022}. The Mexican government announced stringent physical distancing policies on March 30th, 2020 which resulted in marked changes in the mobility network (Fig.~\ref{fig:fig2}a, Table \ref{tab:s1}).

To quantify the degree to which mobility patterns are structured by geopolitical boundaries, we use a community detection algorithm that groups municipalities based on their movement patterns \cite{rosvall_maps_2008}. Specifically, we aim to identify groups of municipalities such that movements between municipalities within the same group, i.e., community, are more frequent than movements to other municipalities in other communities. Community detection is often accomplished via modularity maximization \cite{newman_modularity_2006}; however, these approaches neglect information about the flow of mobility through the network. Instead, we leverage the \textit{map equation} via an algorithm called InfoMap \cite{rosvall_maps_2008}. The InfoMap algorithm utilises an information theoretic approach to derive expected connectivity patterns if the observed flows were entirely determined by a random walk process. For this study, InfoMap is ideal because it is conceptually related to infectious disease transmission models, which often also utilise stochastic processes \cite{pei_burden_2021}.

The aim is to identify municipalities where frequent interactions between individuals occur, such that the detected communities approximate the spatial scales of disease transmission (i.e., communities in which it is assumed that infection spreads via contacts within a relatively homogeneously mixing population \cite{clauset_hierarchical_2008}).  Accounting for spatial heterogeneity is known to be important for assessing strategies for interventions \cite{may_spatial_1984}, especially in areas that have marked differences in urban and rural areas \cite{galindoperez_territorial_2022}. Using this algorithm, we identify 16 communities before the first cases of COVID-19 were detected in Mexico (Fig.~\ref{fig:fig3}b). Community size and organisation changed following the announcement of the lockdown (March 23 and 30, 2020) in Mexico and communities generally became smaller (fewer municipalities within each community (Extended Data Figures \ref{fig:SI_allmaps} and \ref{fig:SI_ncomms} show the communities for each week during the study period). At the peak of the lockdown, we identified approximately 60 movement communities (a 4-fold increase from the baseline period).

More specifically, there are two notable shifts in the network following the introduction of NPIs. First, more communities are identified but importantly the size of these communities shrinks disproportionately so that one community expands (Mexico City) and many very small ones emerge (Fig.~\ref{fig:fig2}d). Further, as a result of the lockdown human movements across municipalities decline more rapidly than movements within a community with one important exception: Mexico City. There we observe that the ratio of within municipality movements declines at a similar rate than movements across municipalities (Extended Data Figure \ref{fig:SI_withinbetween}) further proving its central importance in the mobility network in Mexico.

We then compared the weekly infection incidence growth rates within each community and contrasted them to growth rates under a scenario in which municipalities are grouped based on state boundaries (black lines, Fig.~\ref{fig:fig3}a,b). As expected, we find that epidemics in municipalities that are grouped by human mobility were more synchronised compared to those grouped by state (Fig.~\ref{fig:fig3}c; see Extended Data Figure \ref{fig:SI_intuition} for an illustration of the variance in municipality epidemic growth rates for several example groups of municipalities defined by administrative or network boundaries). The synchrony among municipalities within each community were maximised in April and May 2020, a period when cases were rapidly rising across the country. After June, epidemics that are grouped by movement are still more synchronised, but the differences with groupings by state appear to be smaller (Fig.~\ref{fig:fig3}c). This later period (June to October 2020) is a time when Mexico City appears to also lose importance in seeding the epidemic across the country, and local factors (e.g., population size) become more important in determining the epidemic trajectory \cite{kishore_evaluating_2022}. These results are expected as local factors become more influential in determining disease dynamics (population size, local mixing) and that the importance of continued virus re-importations wanes through time \cite{kraemer_spatiotemporal_2021}.

\section{Discussion \& Limitations}\label{sec:discussion}
We present a generalisable approach for understanding the spatial structure of transmission of COVID-19 and other emerging infectious diseases by accounting for the variations of the human mobility network. We aimed to differentiate the transmission dynamics at a level defined by administrative boundaries from that defined by simple community detection algorithms that are applied to aggregated anonymized weekly human mobility data. We find that as human mobility network structures change, so does to spatial transmission. Incorporating these findings into real-world public health decision-making may result in more effective strategies to control an epidemic \cite{ruktanonchai_assessing_2020, JRC122800, joint_research_centre_european_commission_mapping_2020, de_anda_jauregui_modular_2022}. The European Commission for example published a report on Mobility Functional Areas (MFAs) which were informed by mobile phone data but the adoption of these recommendations remained sparse \cite{JRC122800}.

Our model and results are only as accurate as the data that go into them. The Mexican COVID-19 database may suffer from underreporting due to testing shortages, changing case definitions and spatial heterogeneity in reporting \cite{agren_mexico_2020, exceso_mortalidad}. For example, relatively few cases were reported from Oaxaca (Fig.~\ref{fig:fig1}a) which may be due to barriers to access to testing \cite{foundation_mexicos_nodate}. Future extensions of the model and as the pandemic continues will need to take into account high-resolution SARS-CoV-2 cross-immunity. Further, our model is based on higher level descriptions of the population (raw case data and population level human movement data) and these do not capture the high contact heterogeneity within each municipality (e.g., demographic heterogeneity and assortative mixing) shown to be important in the transmission of COVID-19 \cite{chang_mobility_2021}. Contact patterns may differ significantly by age group, employment status and other factors not accounted for in this work. We did however observe heterogeneity in the demographic makeup of cases during the earlier phases of the Mexican COVID-19 pandemic.

Further, results should be interpreted in light of important limitations related to the human mobility data. First, the Google mobility data is limited to smartphone users who have opted into Google’s \textit{Location History} feature, which is off by default. These data may not be representative of the population as whole, and furthermore their representativeness may vary by municipality. Importantly, these limited data are only viewed through the lens of differential privacy algorithms, specifically designed to protect user anonymity and obscure fine detail.

Mexico is composed of 31 free and sovereign states and Mexico City, united under a federation. This means that each administrative region or state is governed by its own constitution, although they are not completely independent of the federal jurisdiction. Furthermore, each state is divided into municipalities, the nation’s basic administrative unit, which possesses limited autonomy (discretionary power on how best to respond to, or apply a public policy). Under a serious nationwide health threat or emergency, such as a pandemic, the federal Ministry of Health (MoH) acquires full authority over the health policies to be implemented nationwide. Nevertheless, Mexican law establishes that the General Health Council (GHC), a collegial body that reports to the president of the republic has the character of health authority, and can emit obligatory norms to be abided by the MoH. The GHC is presided by the Minister of Health, and is conformed by federal institutions (e.g.h, Economy, Communication \& Transport) as well as academic institutions, representatives from pharmaceutical industry, and other health system actors \cite{consejo_website}. Given its mandate and position in the Mexican health system, the GHC constitutes a promising agent to drive public policy outside of the margins or across geo-administrative units. Furthermore, there are examples of inter-state and inter-municipality coordination to resolve problems that extend beyond their borders such as waste management, tax, policing, and perhaps most relevant, health provision. It is in these contexts where evidence-based interventions on innovative approaches, such as the ones presented here become not only an option but a possibility, with greater impact in reducing transmission as compared to approaches where interventions are based on administrative boundaries. 
However, theory often differs from practice and reality brings along additional and expected factors into play (e.g., economic \cite{oliu_barton_effect_2022} and political interests) many of which are not accounted for in this work. Some state governors for example refused to comply with federal health policies in the early relaxation phase in May 2020 \cite{noauthor_lopez_gatell_2020}.

Mexico has suffered a large and devastating epidemic, and we hope that our findings contribute to a more rational implementation of interventions in the future that can account for the substantial and changing spatial heterogeneity in transmission. Such analyses can be updated and translated to any other country in the world for which aggregated human mobility data is available. Future work should also focus on validating the inferred spatial scales with genomic data \cite{hill_progress_2021, kraemer_spatiotemporal_2021, mccrone_contextspecific_2022} or other coarse-graining techniques \cite{klein_emergence_2020, peixoto_bayesian_2019}. Developing interventions using patterns observed in empirical mobility networks must be added to the list of priorities for pandemic response and preparedness in the 21st century.

\section{Materials \& Methods}\label{sec:methods}

\paragraph{Epidemiological data:} Epidemiological data include individual level information on patients with confirmed RTq-PCR COVID-19 infection between March - September 30th, 2020. Data were downloaded from \url{http://datosabiertos.salud.gob.mx/gobmx/salud/datos\_abiertos/datos\_abiertos\_covid19.zip} (last accessed October 24, 2020). Data include information about patients demographics (age and sex) and municipality of residence. In all analyses we used the date of onset of symptoms.

\paragraph{Population and travel data:} Human mobility and population data were extracted at the municipality level based on the 2016 boundaries (INEGI 2016: \url{https://www.inegi.org.mx/app/mapa/espacioydatos/default.aspx}). Population data were downloaded from the COVID-19 indicator dataset, which was provided by INEGI (\url{https://www.inegi.org.mx/investigacion/covid/}).

\paragraph{Aggregated and anonymised human mobility data:} We used the Google COVID-19 Aggregated Mobility Research Dataset described in detail in \cite{kraemer_mapping_2020, lemey_untangling_2021}, which contains anonymized relative mobility flows aggregated over users who have turned on the \textit{Location History} setting, which is turned off by default. This is similar to the data used to show how busy certain types of places are in Google Maps---helping identify when a local business tends to be the most crowded. The mobility flux is aggregated per week, between pairs of approximately 5km$^2$ cells worldwide, and for the purpose of this study further aggregated for municipalities in Mexico.

To produce this dataset, machine learning is applied to log data to automatically segment it into semantic trips. To provide strong privacy guarantees \cite{wilson_differentially_2019}, all trips were anonymized and aggregated using a differentially private mechanism to aggregate flows over time (see \url{https://policies.google.com/technologies/anonymization}). This research is done on the resulting heavily aggregated and differentially private data. No individual user data was ever manually inspected, only heavily aggregated flows of large populations were handled. All anonymized trips are processed in aggregate to extract their origin and destination location and time. For example, if $n$ users travelled from location $a$ to location $b$ within time interval $t$, the corresponding cell $(a,b,t)$ in the tensor would be $n \pm err$, where $err$ is Laplacian noise. The automated Laplace mechanism adds random noise drawn from a zero mean Laplacian distribution and yields $(\epsilon, \delta)$-differential privacy guarantee of $\epsilon = 0.66$ and $\delta = 2.1 \times 10^{29}$ per metric. Specifically, for each week $W$ and each location pair $(A,B)$, we compute the number of unique users who took a trip from location $A$ to location $B$ during week $W$. To each of these metrics, we add Laplace noise from a zero-mean distribution of scale $1/0.66$. We then remove all metrics for which the noisy number of users is lower than 100, following the process described in \cite{wilson_differentially_2019}, and publish the rest. This yields that each metric we publish satisfies $(\epsilon,\delta)$-differential privacy with values defined above. The parameter $\epsilon$ controls the noise intensity in terms of its variance, while $\delta$ represents the deviation from pure $\epsilon$-privacy. The closer they are to zero, the stronger the privacy guarantees.

These results should be interpreted in light of several important limitations. First, the Google mobility data is limited to smartphone users who have opted into Google’s \textit{Location History} feature, which is off by default. These data may not be representative of the population as whole, and furthermore their representativeness may vary by location. Importantly, these limited data are only viewed through the lens of differential privacy algorithms, specifically designed to protect user anonymity and obscure fine detail. Moreover, comparisons across rather than within locations are only descriptive since these regions can differ in substantial ways.

\paragraph{Timeline of interventions:} The Mexican government has outlined four principle objectives for the control of COVID-19: a) Reduce risk of acquiring infection, b) Reduce risk of severe morbidity and mortality, c) Reduce risk and impact on society and d) Reduce risk of transmission between infectious and susceptible individuals. We collated a full list of interventions between February and September 2020 and details are provided in Table \ref{tab:s1}, including references.

\paragraph{Relative risk model:} Following Goldstein and Lipsitch \cite{goldstein_temporal_2020} we used age stratified epidemiological data to assess the temporal shifts in the share of a given age group among all cases of infection. To do so we use the relative risk (RR) \cite{goldstein_relative_2018, goldstein_temporally_2017} statistic that estimates the ratio of the proportion of a given age group among all detected cases of COVID-19 for a later time period vs. an early time period. We selected the early time period to be the month of April (the period right after the implementation of the lockdown) and the late period to be June to September. We adopted the code and model from Goldstein and Lipsitch described in detail \cite{goldstein_temporal_2020}.
 
\paragraph{Community detection algorithm:} Human mobility networks, based on data from mobile devices, can be used to capture important population-level trends. Microscopic descriptions often remain too complex to extract meaningful information to describe the transmission process accurately \cite{chang_mobility_2021}. We here use a community detection algorithm following \cite{rosvall_maps_2008} to identify human movement communities (basins) where within-community mobility among municipalities is higher than across-community mobility. We chose this community detection algorithm as it is conceptually related to infectious disease transmission models---both utilising random walks.

\paragraph{Municipality level case growth rates:} To estimate the daily epidemic growth rates in each municipality, we fit a mixed effects GLM of log new daily case counts in sliding 7-day windows (fixed effect; approximately the generation time of COVID-19 in the earliest wave) and a random effect for each municipality on the slope and intercept, using the R package \texttt{lme4} v.1.1-21 \cite{ripley_mass_2022}. Daily case counts were determined using the date of symptom onset.

\paragraph{Relationship between case growth rates and mobility:} To test for an effect of mobility from Mexico City on municipality growth rates, we fit a mixed effect GLM with log mobility as a fixed effect, a random effect on the intercept for each municipality and a random effect on the slope and intercept for log mobility each week. The conditional and marginal coefficient of determination, i.e., $R^2$, were calculated using the R package MuMIn v1.471. \cite{MuMIn1471} which implements the method developed by Nakagawa et al. 2017 \cite{nakagawa2017coefficient}. Model selection was performed using analysis of variance for mixed effects models as implemented in the R package lmerTest v.3.1-3 \cite{lmerTest313}.

\section*{Additional information}
\paragraph{Acknowledgments:} We thank all health care workers and those involved in the collection, processing and publishing COVID-19 epidemiological data from Mexico. \paragraph{Funding:} M.U.G.K., O.G.P., B.G.~acknowledge funding from the Oxford Martin School Pandemic Genomics programme. M.U.G.K.~acknowledges funding from the European Horizon 2020 programme MOOD (grant no.~\#874850), the Wellcome Trust, a Branco Weiss Fellowship, The Rockefeller Foundation and Google.org. The contents of this publication are the sole responsibility of the authors and do not necessarily reflect the views of the European Commission or the other funders. B.K., H.H., S.V.S., \& A.V.~acknowledge the support of a grant from the John Templeton Foundation (61780). The opinions expressed in this publication are those of the author(s) and do not necessarily reflect the views of the John Templeton Foundation. \paragraph{Author contributions:} S.V.S., M.U.G.K. and B.K.~developed the idea, planned the research and conducted analyses. A.C.Z.~and D.B.S.C.~collected government intervention data. S.V.S., M.U.G.K.~and B.K.~wrote the first draft of the manuscript. All authors interpreted the data, contributed to writing and approved the manuscript. \paragraph{Competing interests:} We declare no conflicts of interest. \paragraph{Data and materials availability:} Code, spatial, and epidemiological data are available upon publication. The Google COVID-19 Aggregated Mobility Research Dataset used for this study is available with permission from Google LLC. Correspondence and requests for materials should be addressed to B.K., J.A.D-Q., S.V.S., or M.U.G.K.

\begin{sloppypar}
\printbibliography[title={References}]
\end{sloppypar}

\clearpage

\appendix
\setcounter{figure}{0}
\setcounter{table}{0}
\setcounter{equation}{0}
\renewcommand\thefigure{\thesection.\arabic{figure}}
\renewcommand\thetable{\thesection.\arabic{table}}
\renewcommand\theequation{\thesection .\arabic{equation}}
\begin{refsection}

\section{Extended Data Figures}\label{sec:appendix_A}

\begin{figure}[h]
    \centering
    \includegraphics[width=1.0\columnwidth]{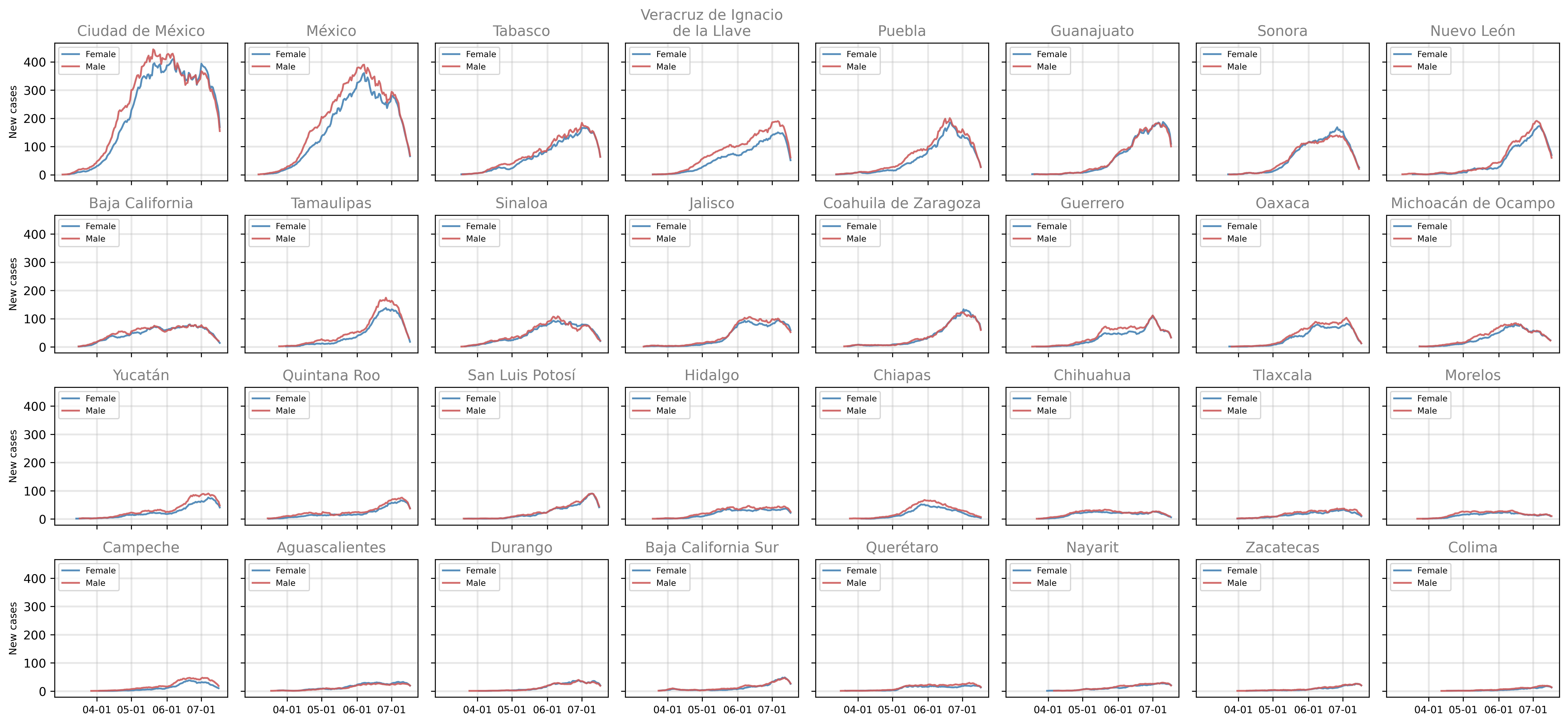}
    \caption{Number of new cases per state and sex (7-day average).}
    \label{fig:SI_region_cases}
\end{figure}
\clearpage

\begin{figure}[h]
    \centering
    \includegraphics[width=1.0\columnwidth]{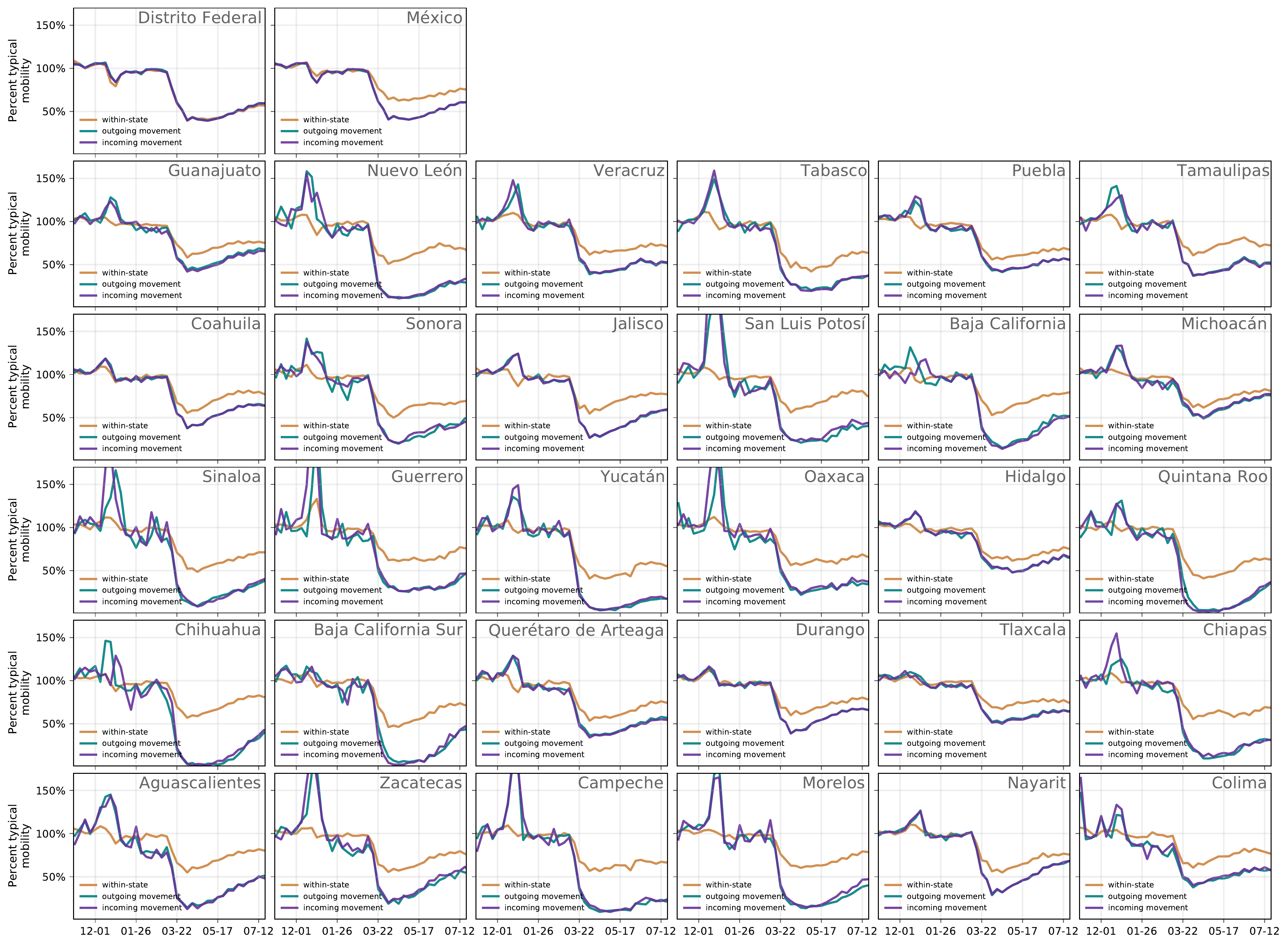}
    \caption{Weekly relative change in human mobility within each state and between states (incoming and outgoing) as compared to baseline.}
    \label{fig:SI_withinbetween}
\end{figure}
\clearpage

\begin{figure}[h]
    \centering
    \includegraphics[width=1.0\columnwidth]{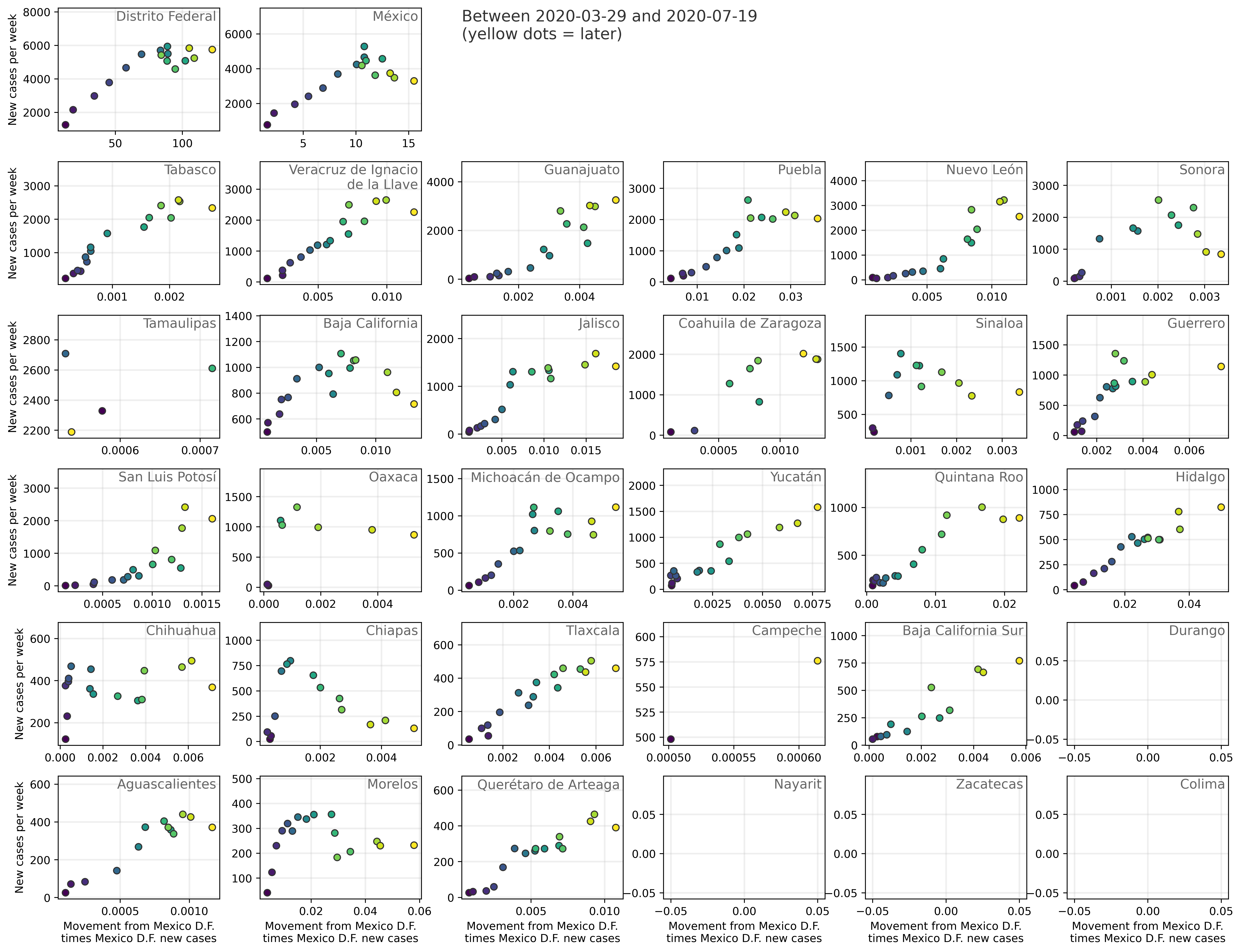}
    \caption{State-specific correlations of new reported cases (weekly) vs.~mobility from Mexico City \textit{times} new reported cases in Mexico City (weekly). States with low mobility and case count data coverage are included but not plotted in this figure.}
    \label{fig:SI_from_to_mxdf}
\end{figure}
\clearpage

\begin{figure}[h]
    \centering
    \includegraphics[width=1.0\columnwidth]{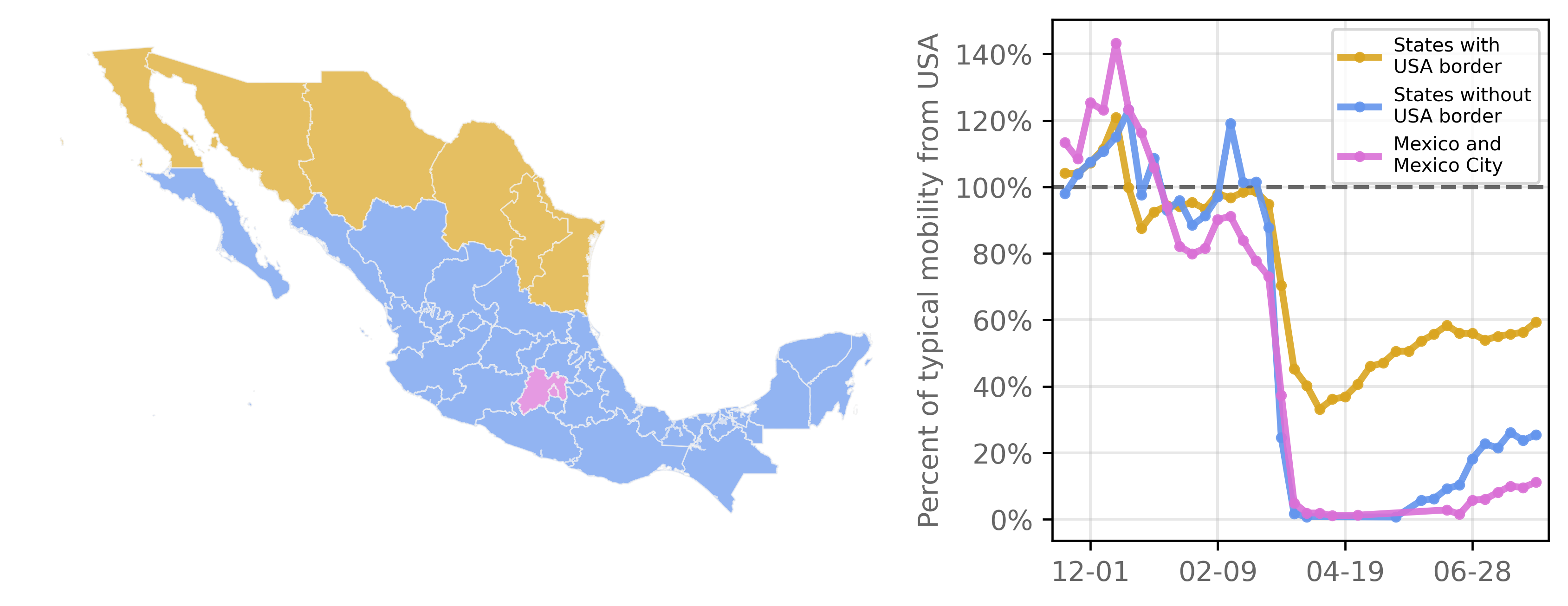}
    \caption{Weekly relative human mobility where the origin is the USA and the destination are states in Mexico divided into states that share a land border, Mexico and Mexico City and all other states.}
    \label{fig:SI_border_mobility}
\end{figure}
\clearpage

\begin{figure}[h]
    \centering
    \includegraphics[width=1.0\columnwidth]{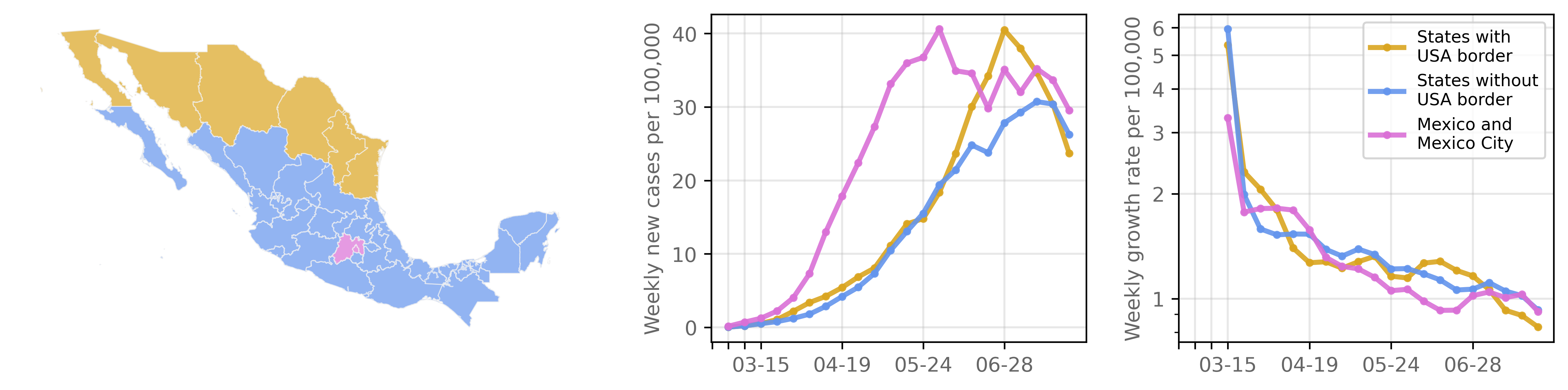}
    \caption{Weekly new cases per 100,000 divided into cases in Mexico City and the state of Mexico, states that share a land border with the USA, and all other states.}
    \label{fig:SI_border_cases}
\end{figure}
\clearpage

\begin{figure}[h]
    \centering
    \includegraphics[width=1.0\columnwidth]{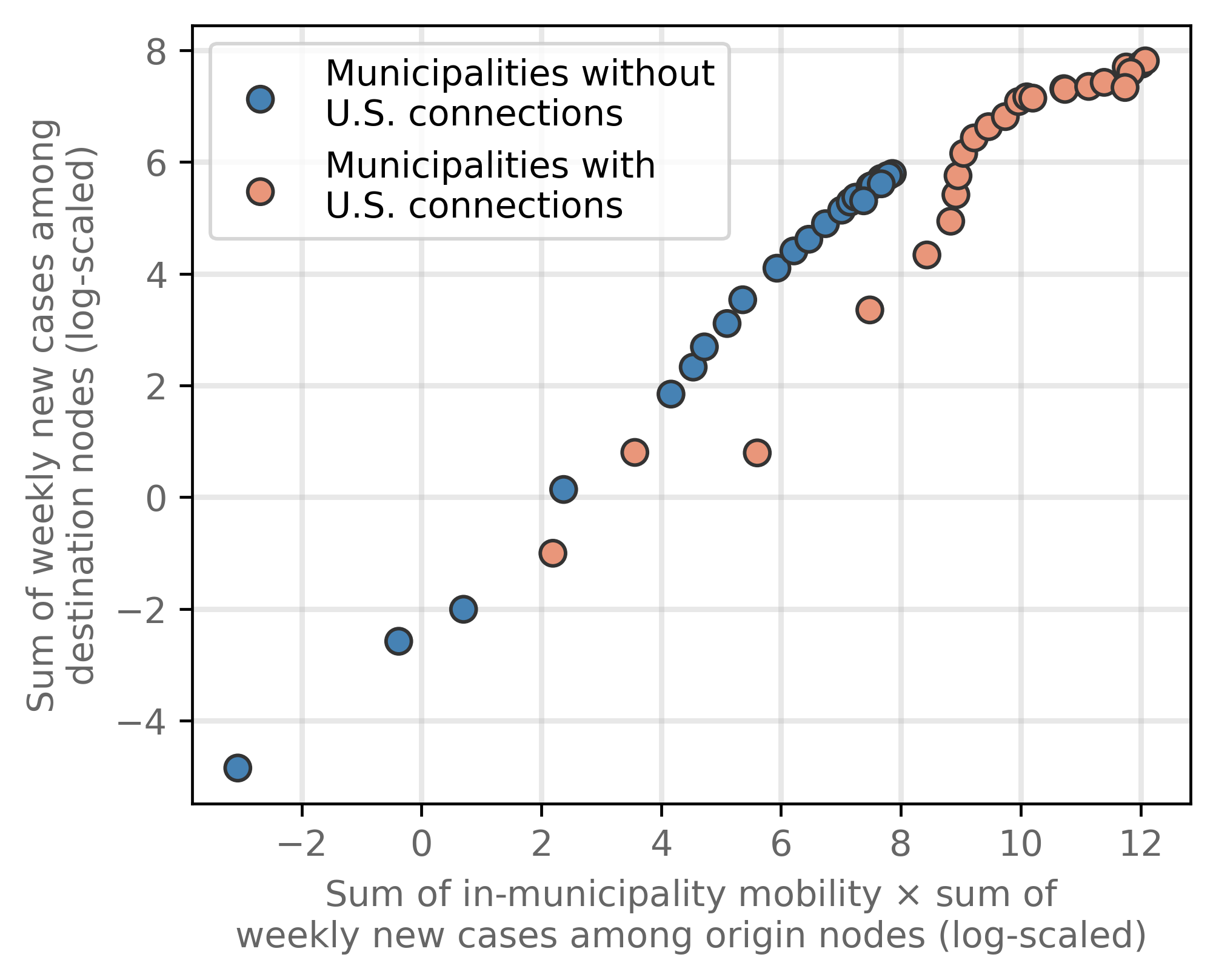}
    \caption{Weekly number of cases among municipalities in Mexico coloured by their geographic position to the USA (bordering vs.~not bordering) and the sum of in-municipality mobility $\times$ weekly new cases among origin nodes (both on the log scale).}
    \label{fig:SI_border_cases_mobility}
\end{figure}
\clearpage

\begin{figure}[h]
    \centering
    \includegraphics[width=1.0\columnwidth]{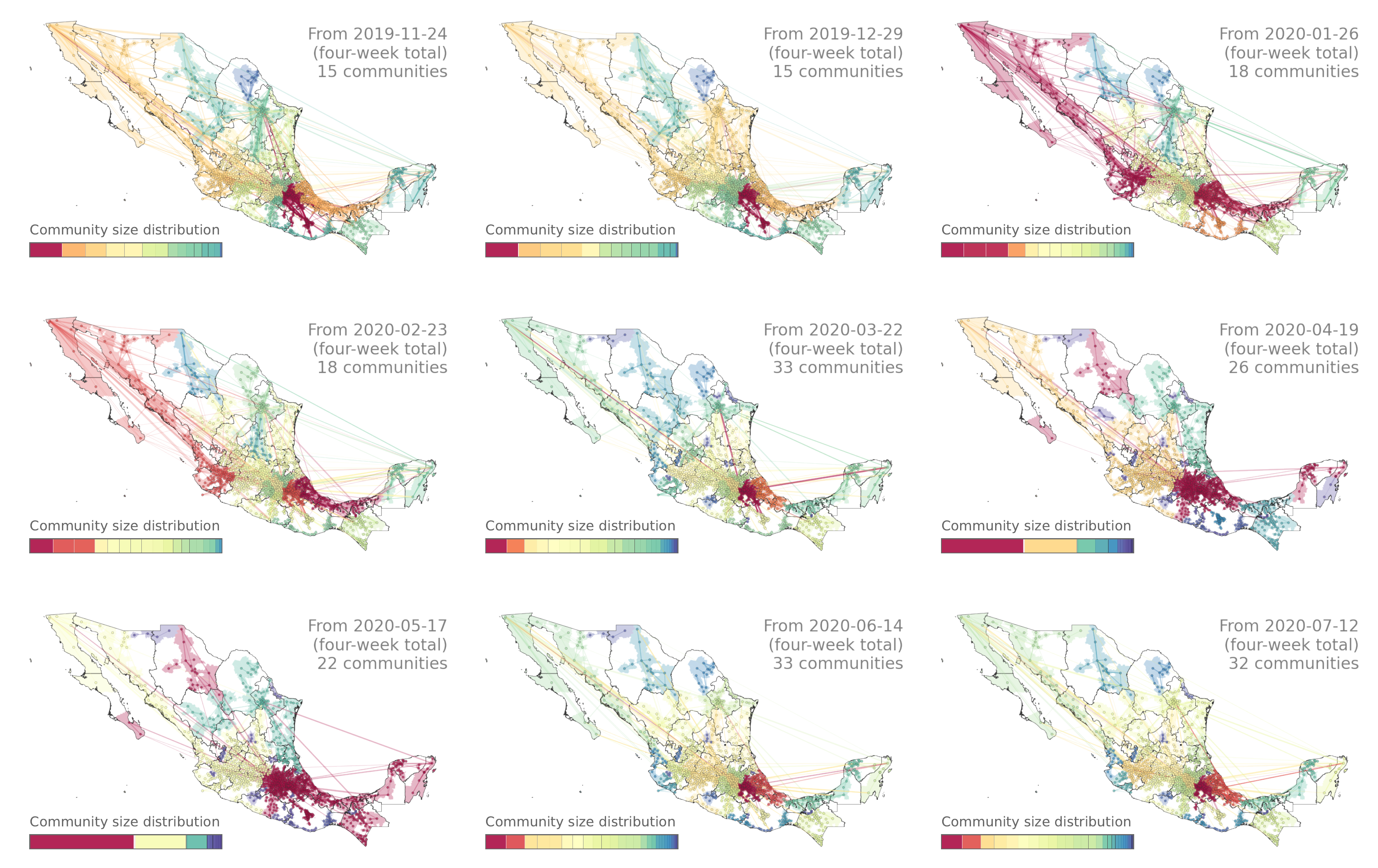}
    \caption{\textbf{Four-week snapshots of mobility in Mexico.} Weekly human mobility in Mexico at the municipality level. Thickness of lines represents intensity of relative mobility flow. Colours represent the membership to movement communities as estimated using the map equation (Materials \& Methods).}
    \label{fig:SI_allmaps}
\end{figure}
\clearpage

\begin{figure}[h]
    \centering
    \includegraphics[width=1.0\columnwidth]{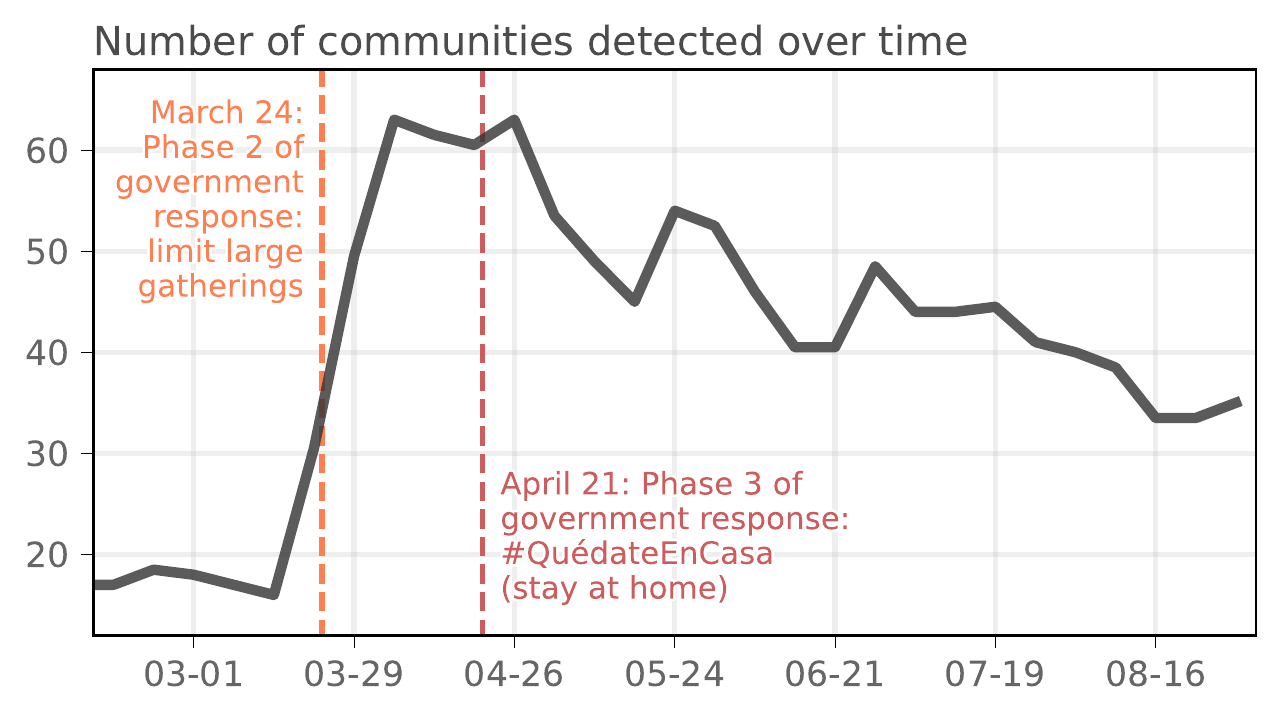}
    \caption{Number of communities detected each week during the first wave of the COVID-19 epidemic in Mexico.}
    \label{fig:SI_ncomms}
\end{figure}
\clearpage

\begin{figure}[h]
    \centering
    \includegraphics[width=1.0\columnwidth]{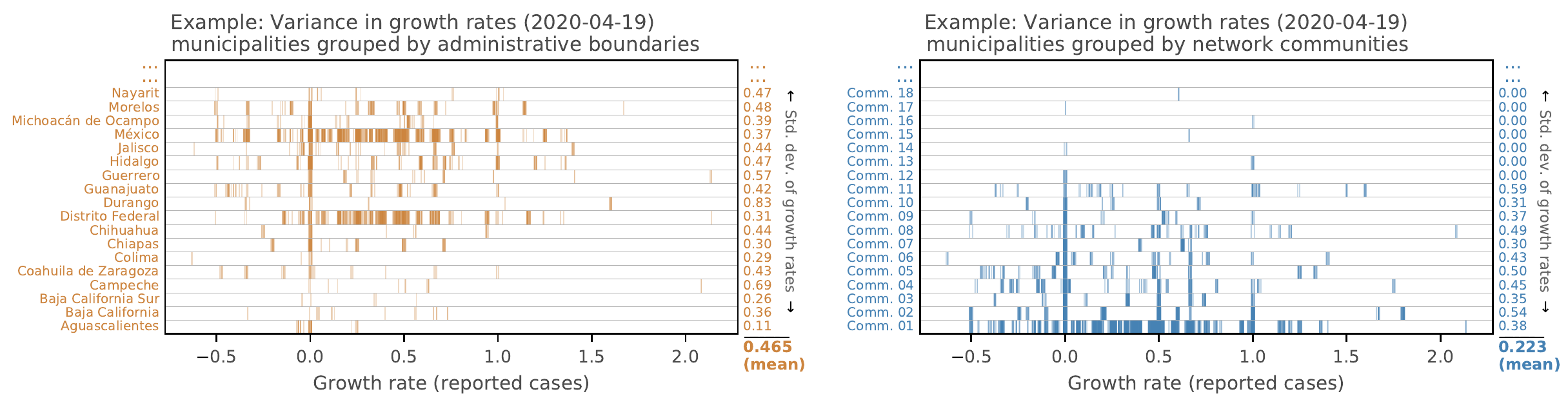}
    \caption{For one example week (April 19, 2020), comparison of the mean standard deviations in municipality growth rates within either (left) administrative (states) boundaries or (right) network communities. In each panel, the standard deviations of municipalities' infection growth rates within each grouping (state vs.~network community) is shown on the right. Figure \ref{fig:fig3}c shows the average of these values over time.}
    \label{fig:SI_intuition}
\end{figure}

\clearpage

\rowcolors{2}{gray!15}{white}
{\renewcommand{\arraystretch}{1.5}
\begin{longtable}{|p{3.5cm}|p{12.0cm}|}
    \hline
    \rowcolor{gray!70}
    \textbf{Date} & \textbf{Intervention} \\ \hline
    March 16, 2020 & Mexican Secretariat of Public Education (SEP) suspend classes in schools of preschool, primary, secondary education, as well as those of the upper middle and higher types dependent on the SEP \cite{noauthor_dof_mar16}. \\ \hline
    March 17, 2020 & Universities begin to suspend classes and social events \cite{march_12_coronavirus_nodate}. \\ \hline
    March 20, 2020 & Mexican Secretariat of Public Education (SEP) cancels all civic and sports events \cite{facebook_gobierno_2020}. \\ \hline
    March 21, 2020 & United States - Mexico border was closed to non-essential travel but remained open for commerce. Closure extended until November 21, 2020 \cite{travel_restrictions}. \\ \hline
    March 23/24, 2020 & National period of social distancing begins. Schools closed and all non-essential operations were closed including gatherings of 100+ people \cite{noauthor_inicia_nodate}. \\ \hline
    March 30, 2020 & National health emergency declared. Policies included: (1.) Non-essential services suspended. (2.) Private sector is asked to require employees to work from home. (3.) Sectors that kept operating normally: government, health (public and private), public safety, social programs, critical infrastructure, and essential services. A full list of essential services can be viewed here: \url{https://www.dof.gob.mx/nota\_detalle.php?codigo=5590914\&fecha=31/03/2020}. (4.) People over 60 years old are urged to stay home. (5.) Public gatherings of over 50 people are banned. (6.) There was no enforced curfew. Expiration date: April 30, 2020. \\ \hline
    April 5, 2020 & Hospital reconversion strategy guidelines published in order to contain nosocomial transmission \cite{noauthor_state_apr20} \\ \hline
    April 16, 2020 & The Federal Government announces the extension of the health emergency and emphasizes the need to restrict movement to and from areas of high transmissibility until May 30th \cite{de_2020_coronavirus_nodate}. \\ \hline
    May 14, 2020 & Ministry of Health announces epidemiologic color-coded system to re-open social, educational, economic activities at state level \cite{noauthor_dof_may14}. \\ \hline
    May 18, 2020 & First phase of the ``new normality'' 324 municipalities with no recorded COVID-19 cases are given green light to reopen businesses and schools \cite{noauthor_conferencia_nodate}. Car factories were meant to reopen on June 1st, but began reopening on May 18th under US pressure (Factories remained closed from March 23rd, to May 18th). \\ \hline
    June 1, 2020 & Mexico’s national period of social distancing concludes. A new color-coded system was enacted across the country to assess how quickly states can reopen their economies and schools: red, orange, yellow, and green \cite{noauthor_state_jun20}. \\ \hline
    July 20, 2020 & Daycare centers run by the country’s social security system reopened in coordination with local authority and based on color-coded indicators. Currently, all but 4 states have daycare centers open. \cite{noauthor_imss_nodate} \\ \hline
    October 18, 2020 & The Health Ministry announced that 17 states---Mexico City included---were at alert level orange and 14 were at yellow. Only one state, Campeche, was at green \cite{noauthor_state_oct18}. In states at the orange level, businesses such as hotels and restaurants can reopen while following health protocols such as enforcing limited capacity. Yellow allows for most economic activities to return to normal with some occupancy limits. \\ \hline
    October 22, 2020 & Some local governments have chosen to enact more stringent restrictions than the federal government guidelines, e.g., Jalisco and Chihuahua \cite{noauthor_state_nodate}.\\ \hline
    \caption{Timeline of government interventions in Mexico.}
    \label{tab:s1}
\end{longtable}}

\clearpage \pagebreak
\subsection{Citation diversity statement}\label{sec:citation-div}
Recent work has quantified bias in citation practices across various scientific fields; namely, women and other minority scientists are often cited at a rate that is not proportional to their contributions to the field \cite{Zurn2020, Dworkin2020, Chakravartty2018, Maliniak2013, Dion2018, Caplar2017, Azoulay2020, Ghiasi2018}. In this work, we aim to be proactive about the research we reference in a way that corresponds to the diversity of scholarship in this field. To evaluate gender bias in the references used here, we obtained the gender of the first/last authors of the papers cited here through either 1) the gender pronouns used to refer to them in articles or biographies or 2) if none were available, we used a database of common name-gender combinations across a variety of languages and ethnicities. By this measure (excluding citations to datasets/organizations, citations included in this section, and self-citations to the first/last authors of this manuscript), our references contain 3\% woman(first)-woman(last), 22\% woman-man, 20\% man-woman, 47\% man-man, 0\% nonbinary, 8\% man solo-author, and 0\% woman solo-author. This method is limited in that an author's pronouns may not be consistent across time or environment, and no database of common name-gender pairings is complete or fully accurate.

\begin{sloppypar}
\printbibliography[title={Supplemental References}]
\end{sloppypar}
\end{refsection}

\end{document}